# Fast Computation of Subpath Kernel for Trees

**Daisuke Kimura** DAISUKE_KIMURA@MIST.I.U-TOKYO.AC.JP
**Hisashi Kashima** KASHIMA@MIST.I.U-TOKYO.AC.JP
Graduate School of Information Science and Technology, The University of Tokyo, Hongo 7-3-1, Bunkyo-ku, Tokyo, 113-8656, JAPAN

## Abstract

The kernel method is a popular approach to analyzing structured data such as sequences, trees, and graphs; however, unordered trees have not been investigated extensively. Kimura et al. (2011) proposed a kernel function for unordered trees on the basis of their subpaths, which are vertical substructures of trees responsible for hierarchical information in them. Their kernel exhibits practically good performance in terms of accuracy and speed; however, linear-time computation is not guaranteed theoretically, unlike the case of the other unordered tree kernel proposed by Vishwanathan and Smola (2003). In this paper, we propose a theoretically guaranteed linear-time kernel computation algorithm that is also practically fast, and we present an efficient prediction algorithm whose running time depends only on the size of the input tree. Experimental results show that the proposed algorithms are quite efficient in practice.

## 1. Introduction

### 1.1. Kernels for structured data

Numerous studies on the research in machine learning are concerned with real-valued vectors. However, a considerable part of real world data is represented not as vectors, but as sequences, trees, and graphs. For example, we can represent biological sequences and natural language texts as sequences, parsed texts and semi-structured data, such as HTML and XML, as trees, and chemical compounds as graphs. Extensive studies have been conducted to analyze structured data such as sequences, trees, and graphs, owing to their widespread use in recent years. Among the various existing approaches, a popular approach to such analysis is the kernel method (Schölkopf & Smola, 2002) because it can efficiently work with high-dimensional (possibly, infinite-dimensional) feature vectors if appropriate kernel functions are provided to access data. A framework called the convolution kernel (Haussler, 1999) is widely used for designing kernel functions for structured data, where the structured data are (implicitly) decomposed into substructures, and a kernel function is defined as the sum of kernel functions among the substructures. It is important to design a good kernel function to balance the expressive power of the substructures with efficient algorithms for computing the kernel. In the framework of the convolution kernel, various kernel functions have been proposed for sequences (Lodhi et al., 2002; Leslie et al., 2002), trees (Collins & Duffy, 2001; Kashima & Koyanagi, 2002; Aiolli et al., 2009), and graphs (Kashima et al., 2003; Gärtner et al., 2003).

### 1.2. Tree kernels

In this paper, we focus on tree kernels. The first tree kernel was proposed by Collins and Duffy (2001) for parse trees, and it was then extended to general ordered trees (Kashima & Koyanagi, 2002). More recently, various kernels for ordered trees have been proposed (Moschitti, 2006; Kuboyama et al., 2006; Aiolli et al., 2009; Sun et al., 2011).

Among the existing tree kernels, only a few kernels can handle unordered trees (Fig. 1(a)). In their seminal work, Vishwanathan and Smola (2003) proposed an efficient kernel for unordered trees. Their work is pioneering in two ways: computation can be performed in linear time with respect to the size of two trees by converting the trees into strings. Moreover, in the prediction phase, prediction for a newly coming tree can be made in linear time with respect to the size of the tree. Their kernel employs complete subtrees as features. Figure 1(b) shows all the complete subtrees in the tree shown in Fig. 1(a). More efficient implementa-tion using the enhanced suffix array (ESA) for *strings* was proposed by Teo and Vishwanathan (2006).

More recently, Kimura et al. (2011) proposed another tree kernel using vertical substructures called subpaths. Figure 1(c) shows all the subpaths in the tree shown in Fig. 1(a). Their kernel is useful for capturing vertical substructures responsible for hierarchical information in trees. Note that neither of the complete subtree features and the subpath feature is a superset of the other. Figure 2 shows the experimental comparison of the predictive accuracy of their kernel with that of four other tree kernels using three datasets, including one XML dataset (Zaki & Aggarwal, 2006) and two glycan datasets (Hashimoto et al., 2003; Doubet & Albersheim, 1992)[1]. Note that three kernels were designed by Kashima & Koyanagi (2002), Moschitti (2006) and Aiolli et al. (2009) for ordered trees; hence, we used the order information appearing in the datasets as it is. The results show that the subpath kernel proposed by Kimura et al. (2011) is competitive with the other kernels. Interestingly, the subpath kernel and the kernel proposed by Vishwanathan and Smola (2003) work complementarily.

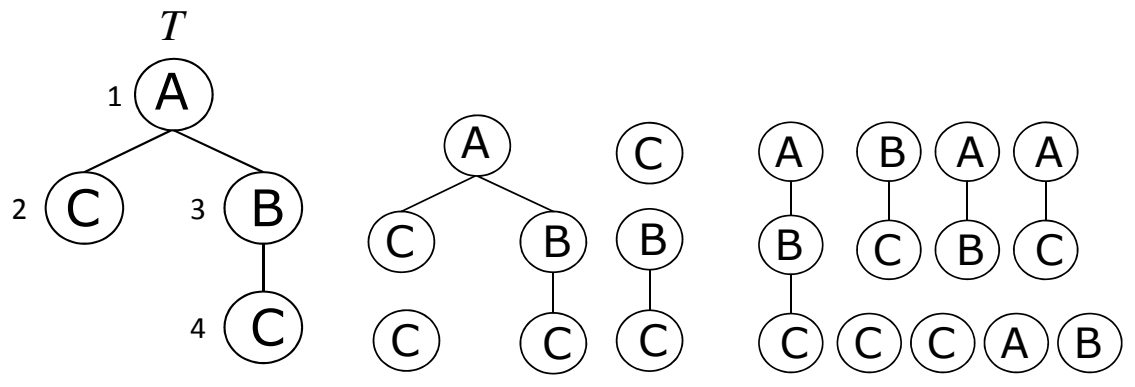

(a) (b) (c)

*Figure 1.* (a) Unordered tree. (b) Complete subtree features of Vishwanathan et al. . (c) Subpath features of Kimura et al. .

Kimura et al. (2011) also showed that their subpath kernel is practically fast and that it is competitive with the linear-time kernel (Teo & Vishwanathan, 2006). However, despite its practical usefulness, the time complexity of the subpath kernel is theoretically $O(n\log n)$ on average, and it is $O(n^2)$ in the worst case, where $n$ is the sum of the sizes of the input trees, because their algorithm for computing the kernel uses the multi-key quick sort (Bentley & Sedgewick, 1997). Moreover, in contrast to the linear-time kernel (Vishwanathan & Smola, 2003; Teo & Vishwanathan, 2006), we need to evaluate the subpath kernel between a given tree and all the support vectors in the prediction phase, which is a serious drawback with large-scale data.

[1] We used LIBSVM (Chang & Lin, 2001) as the SVM implementation. The accuracy is measured using 10-fold cross-validation. Kernels by Kimura et al. (2011) and Moschitti (2006) have tunable weight parameters, which were also tuned by cross-validation. As for datasets, see Table 1.

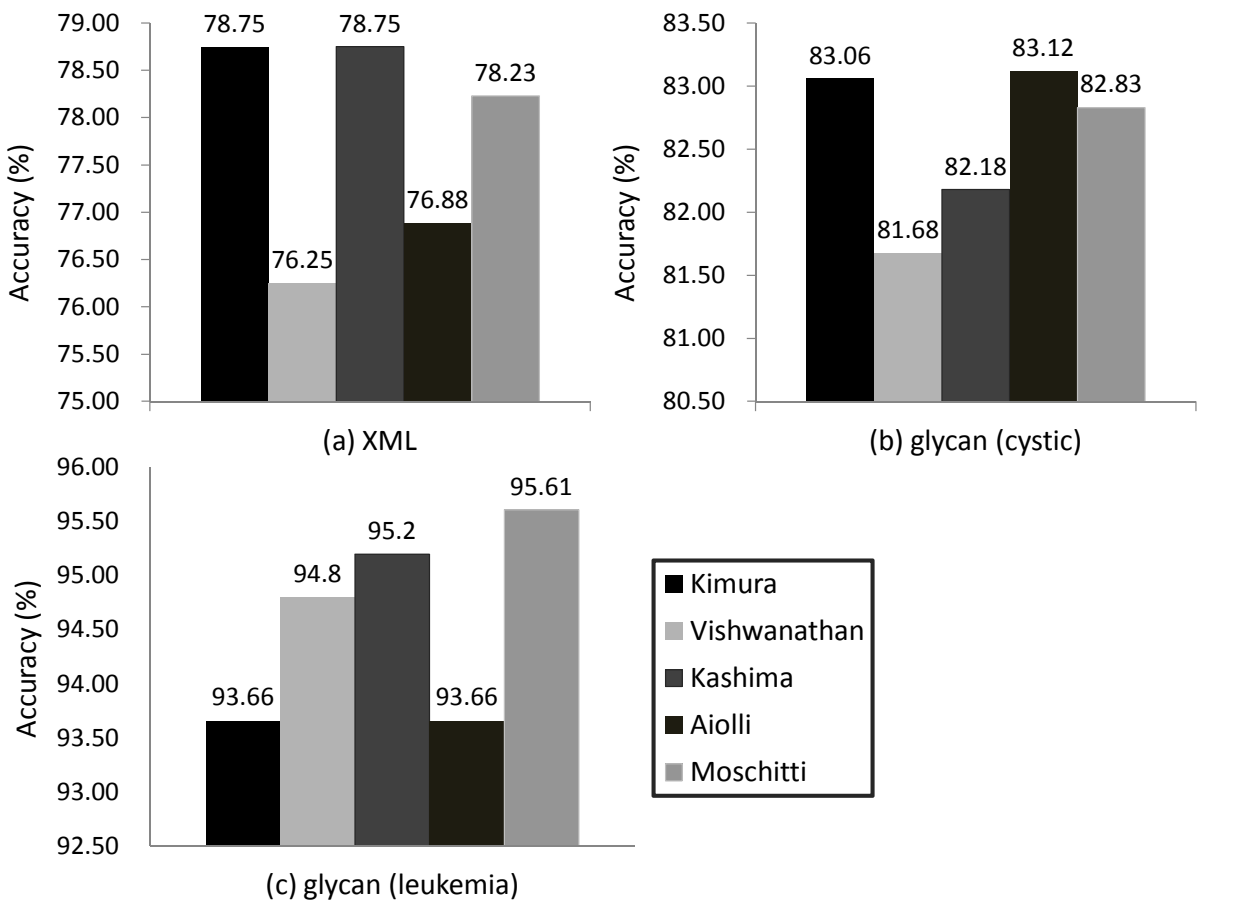

*Figure 2.* Comparison of five tree kernels using three datasets. The subpath kernel proposed by Kimura et al. (2011) is competitive with the three other kernels (Vishwanathan & Smola, 2003; Kashima & Koyanagi, 2002; Moschitti, 2006; Aiolli et al., 2009).

## 1.3. Proposed methods

By improving the result of Kimura et al. (2011), we aim to develop (i) a theoretically guaranteed linear-time kernel computation algorithm that is practically fast, and (ii) an efficient prediction algorithm whose running time depends only on the size of the input tree.

The key to achieving these two objectives is an efficient data structure that accesses vertical substructures in trees. The suffix tree (ST) of trees (Shibuya, 2003) is a potential candidate. However, despite its theoretical merits, its performance (especially its memory usage) is not sufficient for practical use, as pointed out in literature (e.g., Abouelhoda et al. (2004)). In order to overcome this challenge, we use a more space-efficient data structure called the *enhanced suffix array* (ESA) for unordered trees, and we develop its linear-time construction algorithm. To the best our knowledge, this is the first algorithm that constructs the ESA for trees in linear time.

Using the ESA for trees, we devise a linear-time computation algorithm for the subpath tree kernel by simulating bottom-up traversal in the ST. Note that Vishwanathan et al. calculated their kernel by simulating top-down traversal in the ST, using the ESA for *strings*.

We also devise a fast algorithm for prediction, whose complexity is independent of the number of support vectors. This algorithm can be considered as a gener-alization of that of Teo & Vishwanathan (2006), which simulates top-down traversal in the ST. Our algorithm guarantees quadratic time with respect to the size of an input tree in the worst case. Note that a naive implementation using the ESA results in cubic time complexity. Experimental results show the proposed algorithms are also quite efficient in practice.

### 1.4. Contributions

Our study makes the following three contributions:

1. We propose a linear-time algorithm for constructing an enhanced suffix arrays for a tree (Section 3).
2. We propose a linear-time algorithm for computing the subpath kernel (Section 4).
3. We present a fast algorithm for making predictions with the subpath kernel whose time complexity does not depend on the number of support vectors (Section 5).

## 2. Subpath Kernel for Unordered Trees

Kimura et al. (2011) proposed a tree kernel on the basis subpaths to capture vertical substructures responsible for hierarchical information in trees. Formally, a subpath is a substring of a path from the root to one of the leaves (Fig 1(c)). By using subpaths, they proposed a kernel function between two trees $T_1$ and $T_2$ as

$$K(T_1, T_2) \equiv \sum_{p \in P} \lambda^{|p|} \mathrm{num}(T_{1p}) \mathrm{num}(T_{2p}), \qquad (1)$$

where $P$ is the set of all subpaths in $T_1$ and $T_2$, and $\mathrm{num}(T_{1p})$ and $\mathrm{num}(T_{2p})$ are the number of times a subpath $p \in P$ appears in $T_1$ and $T_2$, respectively. $\lambda$ $(0 < \lambda \leq 1)$ is a constant giving an exponentially decaying weight to each subpath $p$, according to its length $|p|$.

Kimura et al. pointed out that there is a one-to-one correspondence between a subpath and a prefix of a suffix of a tree. Intuitively, a prefix of a suffix of a tree is a reversed subpath (whose formal definition is given in Section 3). They utilized this fact, and computed

$$K(T_1, T_2) = \sum_{s \in S} \lambda^{|s|} num(T_{1s}) num(T_{2s}) \qquad (2)$$

by extending the idea of multi-key quicksort instead of computing Eq. (1) directly, where $S$ is the set of all prefixes of suffixes in $T_1$ and $T_2$. The time complexity of their algorithm is $\mathrm{O}((|T_1| + |T_2|)\log(|T_1| + |T_2|))$ on average, and $\mathrm{O}((|T_1| + |T_2|)^2)$ in the worst case.

In Section 4, we will improve this result to $\mathrm{O}(|T_1| + |T_2|)$, even in the worst case, by using the ESA for trees proposed in Section 3.

## 3. Efficient Data Structure for Unordered Trees

First, we review the suffix tree and ESA for a tree. Both of them are essential for constructing fast algorithms for the subpath kernel in Section 4. Then, we propose a novel algorithm for constructing the ESA in linear time.

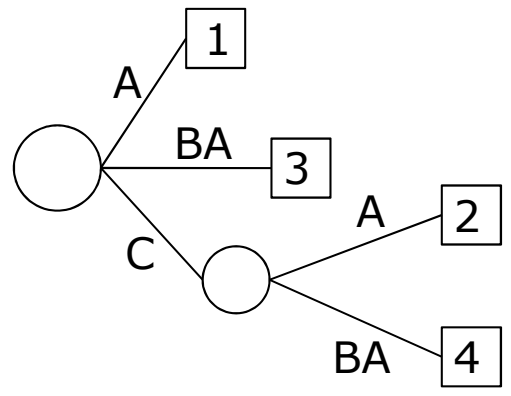

(a)

| *i* | SA | LCP | *Si* |
|---|---|---|---|
| 1 | 1 | 0 | A |
| 2 | 3 | 0 | BA |
| 3 | 2 | 1 | CA |
| 4 | 4 | -1 | CBA |

(b)

*Figure 3.* (a) Suffix tree (ST) for the tree $T$ in Fig 1 (a) maintains all suffixes $S_1, S_2, \ldots, S_4$ of $T$. (b) An ESA for $T$ consists of a suffix array (SA) and an lcp array (LCP).

### 3.1. Suffix Trees and Enhaced Suffix Arrays for a Tree

Let $T$ be a rooted tree consisting of $n$ nodes, whose node labels are drawn from alphabet $\Sigma$ of size $\sigma = |\Sigma|$. The $i$-th *suffix* of $T$ (denoted by $S_i$) is the string associated with the path from the $i$-th node to the root of $T$. For a string $s$, each substring beginning with the first character is called a *prefix*. The *suffix tree* (ST) of a tree $T$ is a patricia trie for all suffixes of $T$ (Fig. 3(a)), where a common prefix of suffixes is associated with a pass from the root to an internal node. Although ST provides fast access to any suffix of the tree, it is known that it requires a large amount of memory, and hence, it is inefficient for practical purposes. The ESA (Abouelhoda et al., 2004) is a more space-efficient data structure that allows many of the operations provided by the ST, and therefore, it is often used in many practical applications instead of the ST.

The ESA of a tree $T$ consists of two data structures, a suffix array (SA) and an lcp array (LCP). The suffix array $\mathrm{SA}[1, |T|]$ is an array of integers that maintains the starting positions of lexicographically ordered suffixes. The lcp array $\mathrm{LCP}[1, |T|]$ is an array of integers that stores the lengths of the longest common prefixes of the adjacent suffixes in the SA, that is, $\mathrm{LCP}[i] \equiv \mathrm{lcp}(S_{\mathrm{SA}[i]}, S_{\mathrm{SA[i+1]}})$ for $1 \leq i < |T|$, and $\mathrm{LCP}[|T|] \equiv -1$, where $\mathrm{lcp}(s, t)$ denotes the longest common prefix of two strings $s$ and $t$. The LCP provides information about the depth of internal nodes of the ST. We present an example of the SA and LCP in Fig 3(b).

### 3.2. Linear-time construction algorithm of an ESA for a tree

There exists a linear-time algorithm that constructs an ESA for a *string* (Kärkkäinen & Sanders, 2003). However, to the best of our knowledge, there is no algorithm for constructing an ESA for a *tree* in linear time. Therefore, we propose an $O(|T|)$ algorithm for constructing an ESA for a tree $T$, which is described as Algorithm 1. Our algorithm is designed by carefully combining two algorithms, namely, the skew algorithm for ESA construction of strings (Kärkkäinen & Sanders, 2003) and the linear-time construction algorithm of an SA for a tree (Ferragina et al., 2005). The following theorem guarantees that this algorithm constructs an ESA of a tree in linear time.

**Theorem 1** *Algorithm 1 constructs the ESA for a tree $T$ in $O(|T|)$ time.*

**Algorithm 1** Constructing the ESA for a tree $T$ in $O(|T|)$ time

**Input:** Tree $T$
**Output:** $SA[1,|T|]$ and $LCP[1,|T|]$ for $T$
1. Apply Algorithm 1 recursively to a tree $T'$ obtain $SA_1[1,|T'|]$ and $LCP_1[1,|T'|]$, where $T'$ consists of some of the nodes in $T$ whose depth is not equal to $d$ in modulo 3
2. Construct $SA_2[1,|T|-|T'|]$ for nodes whose depth is equal to $d$ in modulo 3 by using $SA_1[1,|T'|]$
3. Construct $SA[1,|T|]$ by merging $SA_1[1,|T'|]$ and $SA_2[1,|T|-|T'|]$
4. Construct $LCP[1,|T|]$ by using $LCP_1[1,|T'|]$, $SA_1[1,|T'|]$, and $SA[1,|T|]$
**return** $SA[1,|T|]$ and $LCP[1,|T|]$

In what follows, we give how the algorithm works and a skech of the proof of Thorem 1 by construction.

Algorithm 1 works recursively; it calls itself to construct an ESA for tree $T'$, whose size is at most $\frac{2|T|}{3}$ (in Step 1). We focus only on the suffixes starting at nodes whose depths are not equal to $d \in \{0,1,2\}$ in modulo 3, and apply the radix sorting to them using only the first three labels in each suffix. The parameter $d$ is appropriately chosen so that the number of nodes staring at depth $d$ (in modulo 3) is at least $|T|/3$. Next, we rename the label of each node with the rank of the suffix starting at the node in the sorting result, and construct $T'$ whose size is at most $2|T|/3$. If the all of the renamed node labels are different, their order directly gives the order of suffixes ($SA_1[1,|T'|]$). Otherwise, the algorithm calls itself to construct an ESA for tree $T'$.

In Step 2, it constructs an SA for the nodes not included in the ESA of $T'$. This can be done in $O(|T|)$ time by the radix sort using the first two node labels. (Note that we already know the order of the second node label from $SA_1[1,|T'|]$.) In Step 3, it constructs an SA for $T$ by merging the (E)SA obtained in Step 1 and the SA obtained in Step 2. This can be done in $O(|T|)$ time (Ferragina et al., 2005).

The key to the linear-time construction is in Step 4, which is originated from the algorithm of Kärkkäinen and Sanders (2003) for updating the LCP for strings. Let $k$ and $l$ be $k \equiv SA[i]$ and $l \equiv SA[i+1]$ for $T$, respectively. There are two possible cases: (i) neither of the depths of $k$ and $l$ is equal to $d$ in modulo 3, or (ii) either of them is equal to $d$ in modulo 3.

In the first case, $k$ and $l$ are both included in $T'$. Let $k'$ and $l'$ be the nodes in $T'$ corresponding to $k$ and $l$, respectively. We can know the positions of $k'$ and $l'$ in $SA_1[1,|T'|]$ by using a reversed suffix array (RSA), which is defined as $RSA[SA[i]] \equiv i$. Note that the RSA can also be constructed in linear time. Without loss of generality, we assume that $RSA[k'] < RSA[l']$. Since $k$ and $l$ are adjacent in $SA[1,|T|]$, $k'$ and $l'$ are also adjacent in $SA_1[1,|T'|]$. Since we already have $LCP_1[1,|T'|]$, we obtain $m = \text{lcp}(k',l') = LCP_1[RSA[k']]$ with a constant time by accessing $LCP_1[1,|T'|]$. Then, we can calculate $LCP[i] = 3m + \text{lcp}(\text{anc}(k,3m), \text{anc}(l,3m))$, where $\text{anc}(v_i,j)$ returns the $j$-th ancestor of node $v_i$. Note that three successive node labels in $T$ is renamed with one character in $T'$. In the case with trees, we need preprocessing for all nodes $T$ to achieve $O(1)$-access to their arbitrary ancestor nodes (whereas this is trivial in strings by using the indices). This problem is called the *level ancestor problem.* It can be solved by $O(|T|)$ preprocessing (Bender & Farach-Colton, 2004). Since the $\text{lcp}(\text{anc}(k,3m), \text{anc}(l,3m))$ is at most 2, we can obtain the value in constant time. Consequently, we can compute $LCP[i]$ in $O(1)$ time after $O(|T|)$ preprocessing.

In the latter case, while at least one of $k$ and $l$ is not included in $T'$, $\text{anc}(k,z)$ and $\text{anc}(l,z)$ ($z \in \{1,2\}$) are both included in $T'$. In contrast with the previous case, $k'$ and $l'$ are not necessarily adjacent in $SA_1[1,|T'|]$. However, we can obtain $m = \text{lcp}(k',l') = \text{RMQ}(LCP_1[1,|T'|], RSA[k'], RSA[l'] - 1)$, where $\text{RMQ}(\text{Array},x,y)$ returns the minimum value of $\text{Array}[i]$ ($x \le i \le y$). This problem is called the *range minimum query*, and it can be solved in $O(1)$ time after $O(|T|)$ preprocessing. Consequently, we can compute $LCP[i]$ in $O(1)$ time after $O(|T|)$ preprocessing.

In conclusion, we can compute $\mathrm{LCP}[1, |T|]$ in $\mathrm{O}(|T|)$ time, and hence the total running time of the algorithm represented as $f(T) = f(2|T|/3) + \mathrm{O}(|T|) = \mathrm{O}(|T|)$.

## 4. Linear-time Algorithm for Computing the Subpath Kernel

We propose a linear-time algorithm for computing the subpath kernel (2) by using the ESA for the trees introduced in Section 3. The proposed algorithm consists of the following three steps.

1. Add special terminal characters $\$_1$ and $\$_2$ ($\$_1 < \$_2$) just above the root nodes of $T_1$ and $T_2$, respectively; then, merge the two trees. (We assume that $\$_1$ and $\$_2$ are lexicographically smaller than any character in $\Sigma$.)
2. Construct an ESA for the merged tree by using Algorithm 1.
3. Calculate Eq. (2) by simulating bottom-up traversal in the ST for the merged tree, with the ESA.

First, the algorithm merges the input trees $T_1$ and $T_2$ in Step 1, in which the special terminal characters are added to ensure that no suffix can be a prefix of any other suffix. Next, it constructs the ESA for the merged tree using Algorithm 1 in linear time. Finally, it calculates Eq. (2) with the ESA. Since Eq. (2) can be calculated by enumerating all the common prefixes of suffixes in $T_1$ and $T_2$, it can be computed as

$$K(T_1, T_2) = \sum_{v \in \mathrm{ST_{in}}} (\mathrm{W}[\mathrm{depth}(v)] - \mathrm{W}[\mathrm{depth}(\mathrm{parent}(v))]) \times \mathrm{lvs}(T_{1v})\mathrm{lvs}(T_{2v}),$$

by a bottom-up traversal of the ST for the merged tree. In the above equation, $\mathrm{ST_{in}}$ denotes the set of inner nodes in ST, $\mathrm{depth}(v)$ denotes the depth of node $v$, $\mathrm{parent}(v)$ denotes the parent node of $v$, and $\mathrm{lvs}(T_{1v})$ denotes the number of leaves belonging to $T_1$ among the leaves of the subtree rooted at $v$. W is an array whose elements are defined as $\mathrm{W}[n] \equiv \sum_{i=1}^{n} \lambda^i$, where $\lambda$ is the decaying rate in Eq. (2). Since we constructed an ESA instead of a ST, we use the ESA to simulate bottom-up traversals in the ST (Kasai et al., 2001). Algorithm 2 shows the pseudocode of a bottom-up traversal in the ST and calculation of Eq. (2) with the ESA. Since the number of nodes in the ST for a $T$ is $2|T|$ at most, Algorithm 2 runs in $\mathrm{O}(|T_1| + |T_2|)$ time. Figure 4 shows an example of (a) the merged tree obtained in Step 1, (b) the ST for the merged tree, and (c) the ESA for the merged tree.

The following theorem shows that the three steps listed above compute the subpath kernel (2) in linear time.

**Theorem 2** *The proposed algorithm computes the subpath kernel (2) for $T_1$ and $T_2$ in $\mathrm{O}(|T_1| + |T_2|)$ time.*

**Proof 1** *We can merge $T_1$ and $T_2$ in $\mathrm{O}(1)$ time in Step 1. In Step 2, Algorithm 1 constructs the ESA in $\mathrm{O}(|T_1| + |T_2|)$ time. Finally, Algorithm 2 computes Eq. (2) in $\mathrm{O}(|T_1| + |T_2|)$ time. Therefore, the total time complexity of the proposed algorithm is $\mathrm{O}(|T_1| + |T_2|)$.*

**Algorithm 2** Algorithm for computing Eq. (2) in$\mathrm{O}(|T_1| + |T_2|)$

**Input:** Trees $T_1$ and $T_2$, $\mathrm{SA}[1, |T_1| + |T_2|]$, $\mathrm{LCP}[1, |T_1| + |T_2|]$, Stack $s$, Array W
**Output:** The kernel function value of Eq.(2)
kernel $= 0$, $s = [0, 0, -1, -1]$
**for** $i = 1$ to $|T_1| + |T_2|$ **do**
  **if** $\mathrm{SA}[i] \le |T_1|$ **then**
    $s$.append$(1, 0, i, \mathrm{depth}(\mathrm{SA}[i]))$
  **else**
    $s$.append$(0, 1, i, \mathrm{depth}(\mathrm{SA}[i]))$
  **end if**
  $l_1, l_2, x, h = s.\mathrm{top}()$, $h_i = \mathrm{LCP}[i]$, $c_1 = 0, c_2 = 0$
  **while** $h > h_i$ **do**
    $c_1 += l_1, c_2 += l_2$, $l'_1, l'_2, x', h' = s.\mathrm{top}()$
    $s.\mathrm{pop}(), l_1, l_2, x, h = s.\mathrm{top}()$
    kernel$+= (\mathrm{W}[h'] - \mathrm{W}[h]) * l'_1 * l'_2$
    **if** $h = h_i$ **then**
      $s.\mathrm{top} = (l_1 + c_1, l_1 + c_1, x, h)$
    **else if** $h_i > 0$ && $h < h_i$ **then**
      $s$.append$(c_1, c_2, i, h_i)$
    **end if**
  **end while**
**end for**
**return** kernel

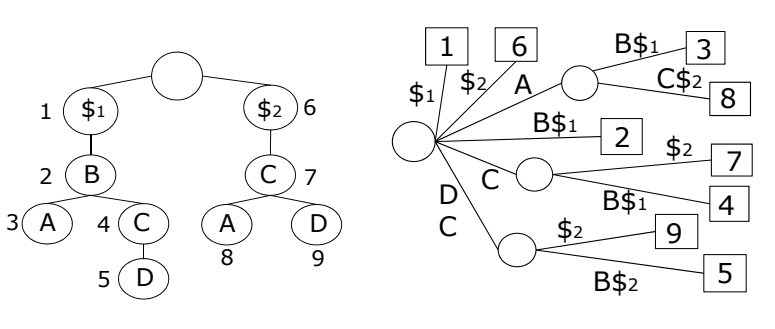

| $i$ | SA | LCP | $S_i$ |
|---|---|---|---|
| 1 | 1 | 0 | $\$_1$ |
| 2 | 6 | 0 | $\$_2$ |
| 3 | 3 | 1 | AB$\$_1$ |
| 4 | 8 | 0 | AC$\$_2$ |
| 5 | 2 | 0 | B$\$_1$ |
| 6 | 7 | 1 | C$\$_2$ |
| 7 | 4 | 0 | CB$\$_1$ |
| 8 | 9 | 2 | DC$\$_2$ |
| 9 | 5 | -1 | DCB$\$_2$ |

(a) (b) (c)

*Figure 4.* (a) The merged tree of two trees. (b) The ST for the merged tree (which we do not explicitly construct). (c) The ESA for the merged tree (with which we simulate the ST).

## 5. Fast Prediction

A serious drawback in applying kernel methods to large-scale data sets is that we need to evaluate a kernel function between an input data $T$ and all the support vectors $T_i$ ($i = 1, ..., m$) in the prediction phase. For the subpath kernel, prediction for a tree $T$ needs

to evaluate

$$f(T) = \sum_{i=1}^{m} \alpha_i K(T_i, T) = \sum_{s' \in PS_i} \sum_{s \in PS} \left( \sum_{i=1}^{m} \alpha_i \mathrm{num}(T_{i\,s}) \right) \lambda^{|s|} \delta(s', s), \quad (3)$$

where $PS_i$ and $PS$ are the set of all prefixes of suffixes in $T_i$ and $T$, respectively. Note that Eq. (3) is identical to the one of Teo and Vishwanathan (2006) when $\{T_i\}_i$ and $T$ are strings. They proposed a sophisticated $\mathrm{O}(|T|)$-time algorithm whose running time does not depend on the number of support vectors. We briefly describe the algorithm. The computation of Eq. (3) comes down to finding the longest common prefix (lcp) between $S_i$ and the "master string", which is the concatenation of all the strings corresponding to the support vectors. First, the algorithm constructs an ESA of the master string, Then, when an input string $T$ comes, the algorithm simulates a top-down traversal of a ST with the ESA to find the lcp between $S_i$ and the master string. The algorithm utilizes the fact that $l_i \geq l_{i-1} - 1$, where $l_i$ is the length of the lcp between $S_i$ and the master string. This implies that we can skip the first $l_{i-1}$ characters and start comparison from the next character when we find $l_i$.

We extend their algorithm to trees. First, we construct an ESA for the "master tree" obtained by concatenating all the support vector trees. Then, when an input tree $T$ comes, the algorithm simulates a top-down traversal of a ST with the ESA to find the lcp between $S_i$ and the master tree. The difference from the previous algorithm is in its skipping strategy. Since node $i$ may have more than one children in trees, $l_i \geq \max l_{Ch(i)} - 1$ holds, where $Ch(i)$ is the set of children of node $i$. Thus, we can skip $\max l_{Ch(i)} - 1$ nodes when we find $l_i$. The following theorem guarantees that the prediction for a newly coming tree $T$ is computed in $\mathrm{O}(|T|^2)$ time in the worst case.

**Theorem 3** *The time complexity of prediction for a tree $T$ is $\mathrm{O}(|T|^2)$.*

**Proof 2** *We evaluate the total length which we traverse the ST to find all the lcp between $S_i$ and the master tree.*

$$\sum_{i:\max l_{Ch(i)}=0} (l_i + 1) + \sum_{i:\max l_{Ch(i)} \neq 0} (l_i - \max l_{Ch(i)} + 2)$$

$$\leq 2|T| + \sum_{i:l_i \neq \max l_{Sib(i)}} l_i$$

$$\leq 2|T| + (\mathrm{L}-1)\mathrm{H} \leq \mathrm{O}(|T|^2).$$

*For any node $i$ satisfying $\max l_{Ch(i)} = 0$, we cannot skip any node; hence, we have to walk down the ST of the master tree (corresponding to the first term in the first line). For any node $i$ satisfying $\max l_{Ch(i)} \neq 0$, we can skip $\max l_{Ch(i)} - 1$ nodes (the second term in the first line). $Sib(i)$ in the right term in the second line denotes the siblings of node $i$. Since the number of nodes in this term is $L - 1$ (where $L$ is the number of leaves in $T$) and $l_i$ is upper bounded by $\mathrm{H}(= \max \mathrm{depth}(T))$, the time complexity is upper bounded by $\mathrm{O}(|T|^2)$.*

Note that the time complexity of the algorithm becomes $\mathrm{O}(|T|)$ for some trees (for example, when either L or H is bounded by a constant). Finally, we point out that the algorithms in Section 4 and this section can also be applied to the route kernel (Aiolli et al., 2009) for *ordered* trees.

## 6. Experiments

We demonstrate the performance of the linear time algorithm for the subpath kernel and the fast computation algorithm in the prediction phase.

First, we compare the execution time of the proposed linear-time algorithm (denoted by 'Proposed') with that of the existing algorithm of Kimura et al. (2011) ('Multikey') for the subpath kernel. We also compare them with the the linear-time tree kernel (Teo & Vishwanathan, 2006) ('Vishwanathan') implemented by Teo and Vishwanathan[2].

Next, we examine the execution time of the algorithm in the prediction phase (denoted by 'Prediction'). We study the effect of the size of an input tree and the number of support vectors. We also compare the execution time with the direct computation of Eq. (3) ('Direct') and the linear-time tree kernel (Teo & Vishwanathan, 2006) ('Vishwanathan').

We run all the experiments on an Intel Core2 Duo 2.40GHz system with 4GB of main memory under Windows Vista. For all the kernels we use in the experiments, we set $\lambda = 1$ in Eq. (2) since the choice does not affect the computation time.

### 6.1. Experiment 1: Fast kernel evaluation

We compare the execution times using three real data sets, including one XML data set (Zaki & Aggarwal, 2006) and two glycan data sets (Hashimoto et al., 2003; Doubet & Albersheim, 1992). Table 1 lists the statistics of these datasets.

We measured the average computation time needed for a single evaluation of each kernel function. Figure 5 (a) shows the average times of 'Proposed', 'Multikey', and 'Vishwanathan' for the three datasets. The results show that our proposed linear-time algorithm is consistently the fastest, which shows that our kernel is quite efficient in practice as well as in theory.

[2] http://users.cecs.anu.edu.au/~chteo/SASK.html

Next, we examine the scalability of the algorithms with artificial datasets. We fixed the label size at 5, and we varied the tree size. Figure 5 (b) shows the average times of the three algorithms. Again, our proposed linear-time algorithm outperforms the others.

*Table 1.* Statistics of real datasets

| data set | data set size | label size |
|---|---|---|
| XML | 3183 | 9097 |
| Cystic | 160 | 27 |
| Leukemia | 480 | 59 |

| data set | avg. nodes | avg. degree | avg. depth |
|---|---|---|---|
| XML | 14.3 | 1.9 | 6.6 |
| Cystic | 8.2 | 1.9 | 4.0 |
| Leukemia | 13.5 | 2.0 | 6.4 |

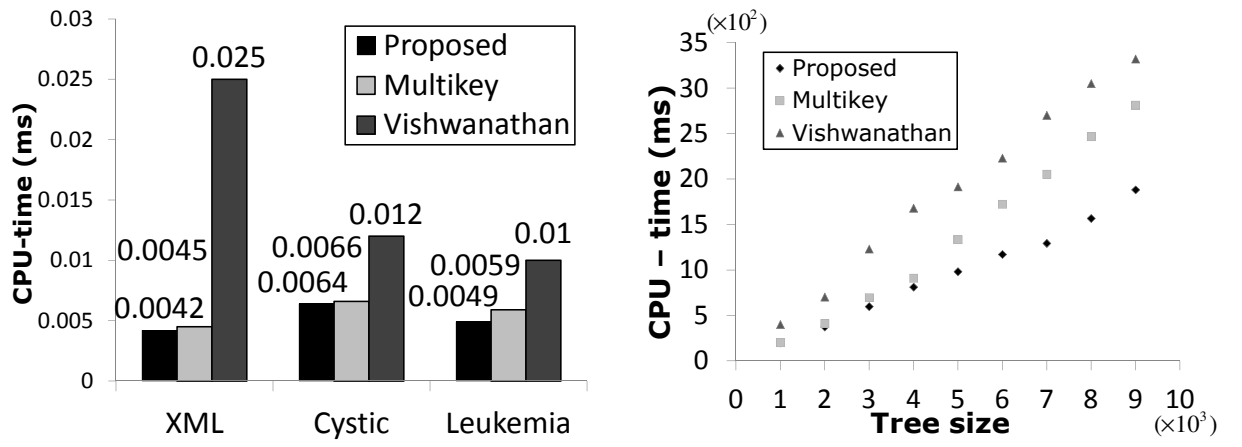

(a) (b)

*Figure 5.* Comparison of execution times for (a) the real datasets and for (b) the artificial datasets.

### 6.2. Experiment 2: Fast prediction

We compare the execution times in the prediction phase with the XML dataset. We give all support vectors a uniform weight of $\alpha_i = 1$.

First, we study the effect of the number of support vectors. We merge the first 100 XML data to make an input tree. We use the other XML data as support vectors, and vary the number of support vectors. Figure 6 (a) shows the average times of 'Prediction', 'Direct', and 'Vishwanathan'. The results show that the execution times of Prediction and Vishwanathan are not dependent on the number of support vectors, whereas that of Direct scales linearly.

Next, we study the effect of the size of an input tree on the 'Prediction' algorithm. We fix the number of support vectors at 100, and we vary the size of an input tree. Figure 6 (b) shows the average times of Prediction. Although the time complexity of the algorithm is theoretically quadratic with respect to the size of an input tree in the worst case, the execution time scales linearly in practice.

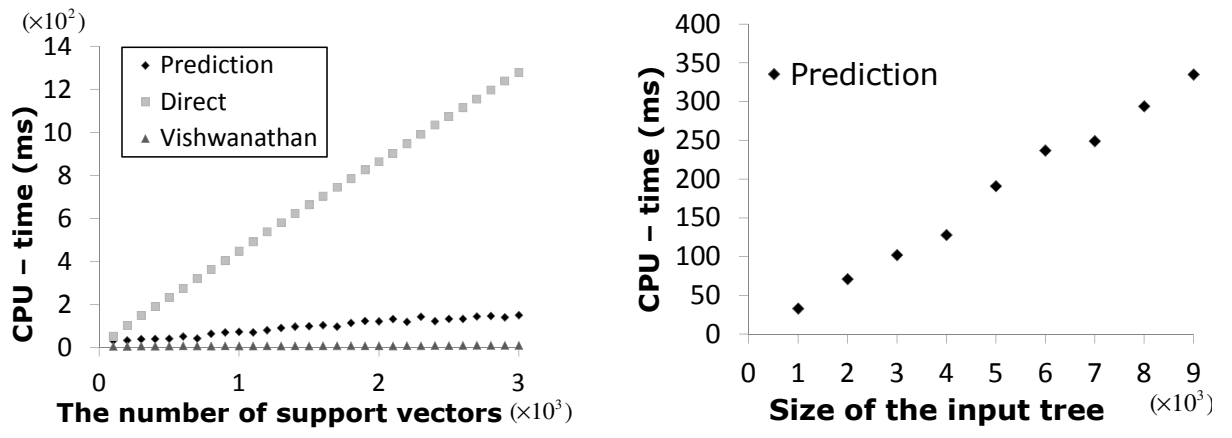

(a) (b)

*Figure 6.* (a)Comparison of execution times in prediction. (b) Execution time for a fixed number of support vectors in prediction.

## 7. Related Work

Since Haussler (1999) introduced the framework of the convolution kernel, various kernel functions for trees have been proposed. The first tree kernel was proposed for parse trees by Collins and Duffy (Collins & Duffy, 2001), and then, it was generalized for labeled ordered trees (Kashima & Koyanagi, 2002; Kuboyama et al., 2006), syntactic trees (Daumé III & Marcu, 2004), and positional trees (Aiolli et al., 2009). However, all these kernels (explicitly or implicitly) exploit edge order information at each node in their definitions or algorithms, and therefore, they cannot be directly applied to unordered trees. For unordered trees, a hardness result for tree kernels using general tree-structured features was shown by Kashima (2007). Vishwanathan et al. (2003) proposed an efficient linear-time kernel based on subtrees. While this kernel can be computed efficiently with the ESA for *strings*, it is pointed out that its predictive performance is usually worse than that of the other tree kernels in the previous work of Aiolli et al. (2009). Kimura et al. (2011) proposed another tree kernel for unordered trees using vertical substructures called subpaths.

## 8. Conclusion

In this paper, we focused on the subpath kernel for unordered trees proposed by Kimura et al. (2011), and we proposed a linear-time algorithm for computing it with an enhanced suffix array for trees. To achieve the desired time complexity, we proposed, for the first time, a linear-time algorithm for constructing an enhanced suffix array for trees. In addition, we presented a fast algorithm for prediction, which is independent of the number of support vectors because it exploits the algorithm of (Teo & Vishwanathan, 2006). Exper-imental results showed that the proposed algorithm is faster than the existing algorithm, and its practical running time scales linearly in practice. Moreover, the running time in prediction is independent of the number of support vectors. A possible future development is to combine the subpath kernel with a fast training framework of SVM with kernels. Recently, Severyn and Moschitti (2011) proposed a fast training algorithm for structured kernels with a cutting plane method, which might be applied for the subpath kernel.

## Acknowledgments

This work was supported by MEXT KAKENHI 80545583.